\documentclass[useAMS,usenatbib,aas_macros]{mn2e}

\usepackage{epsfig}
\usepackage{amssymb}
\usepackage{natbib}
\usepackage{aas_macros}

\begin{document}

\newcommand{\CIV}{C~{\sc iv}}
\def\sarc{$^{\prime\prime}\!\!.$}
\def\arcsec{$^{\prime\prime}\, $}
\def\arcmin{$^{\prime}$}
\def\kms{${\rm km\, s^{-1}}$}
\def\degr{$^{\circ}$}
\def\re{$R_e$}
\def\rb{$R$}
\def\phire{$\Phi(R_e)$}
\def\phirez{$\Phi(R_e,z)$}
\def\phiM{$\Phi(M_{\rm star})$}
\def\phiv{$\Phi(\sigma)$}
\def\sis{$\sigma$}
\def\seco{$^{\rm s}\!\!.$}
\def\ls{\lower 2pt \hbox{$\;\scriptscriptstyle \buildrel<\over\sim\;$}}
\def\gs{\lower 2pt \hbox{$\;\scriptscriptstyle \buildrel>\over\sim\;$}}
\def\mbh{$M_{\rm BH}$}
\def\mstar{$M_{\rm star}$}
\def\msun{${\rm M_{\odot}}$}
\def\MMs{$M_{\rm dyn}/M_{\rm STAR}$}
\def\ML{$M_{\rm dyn}/L_r$}
\def\MsL{$M_{\rm STAR}/L_r$}
\def\MLc{$M_{\rm dyn}/L_{\rm corr}$}
\def\MdM{$M_{\rm dyn}/M_{\rm STAR}$}
\def\Lcorr{$L_r^{\rm corr}$}
\def\lsun{${\rm L_{\odot}}$}
\def\PapI{Shankar et al. (2009a)}
\def\PapInp{Shankar et al. 2009a}

\newcommand{\kpch}{$h^{-1}\,\mbox{kpc}$\,}

\title[]{Further constraining Galaxy evolution models through the Size Function of SDSS Early-type galaxies}

\author[F. Shankar et al.]
{Francesco
Shankar$^{1}$\thanks{E-mail:$\;$shankar@mpa-garching.mpg.de}, Federico
Marulli$^{2}$, Mariangela Bernardi$^{3}$, Michael Boylan-Kolchin$^{1}$, \newauthor Xinyu Dai$^{4}$, and Sadegh Khochfar$^{5}$\\
$1$ Max-Planck-Instit\"{u}t f\"{u}r Astrophysik,
Karl-Schwarzschild-Str. 1, D-85748, Garching, Germany\\
$2$ Dipartimento di Astronomia, Universit\'{a} degli
Studi di Bologna, via Ranzani 1, I-40127 Bologna, Italy\\
$3$ Department of Physics and Astronomy, University of Pennsylvania,
209 South 33rd St,
Philadelphia, PA 19104\\
$4$ Physics and Astronomy Department, University of Oklahoma, Norman, OK, 73019, USA\\
$5$ Max-Planck-Instit\"{u}t f\"{u}r Extraterrestrische Physik, Giessenbachstra{\ss}e, D-85478, Garching, Germany}

\date{}
\pagerange{\pageref{firstpage}--\pageref{lastpage}} \pubyear{2009}
\maketitle

\label{firstpage}


\begin{abstract}
We discuss how the effective radius \re\ function (ERF) recently
worked out by Bernardi et al. (2009) represents a new testbed to improve the current
understanding of Semi-analytic Models of Galaxy formation. In particular, we here show
that a detailed hierarchical model of structure formation can broadly reproduce the correct
peak in the size distribution of local early-type
galaxies, although it significantly overpredicts the number
of very compact and very large galaxies.
This in turn is reflected in the predicted size-mass relation,
much flatter than the observed one, due
to too large ($\gtrsim 3$ kpc) low-mass galaxies ($<10^{11}$\msun),
and to a non-negligible fraction of compact ($\lesssim 0.5-1$ kpc)
and massive galaxies ($\gtrsim 10^{11}$\msun).
We also find that the latter discrepancy is smaller than previously
claimed, and limited to only ultracompact (\re$\lesssim 0.5$ kpc) galaxies
when considering elliptical-dominated samples. We explore several causes
behind these effects. We conclude that the former problem might be linked
to the initial conditions, given that large and low-mass galaxies are present
at all epochs in the model. The survival of compact and massive
galaxies might instead be linked to their very old ages and
peculiar merger histories.
Overall, knowledge of the galactic stellar mass {\em and} size distributions
allows a better understanding of where and how to improve models.
\end{abstract}

\begin{keywords}
galaxies: structure -- galaxies: formation -- galaxies: evolution --
cosmology: theory
\end{keywords}

\section{Introduction}
\label{sec|intro}

The formation and evolution of early-type galaxies, characterized
by a dominant central stellar bulge component, is still a matter of debate.
The seminal paper by
\citet[][]{Eggen62}, postulated that stars are formed in a single burst of
star formation from gas falling towards the center, and the
evolution is passive thereafter. Such a simple
scenario might be difficult to reconcile with the standard cosmological paradigm of structure
formation, in which dark matter halos grow
hierarchically through merging. The most
advanced and up-to-date Semi-analytical models
(SAMs) of galaxy formation \citep[e.g.,][]{Cole00,Benson03,Granato04,Granato06,Menci04,Ciras05,Khochfar05,Vittorini05,Bower06,Cattaneo06,Croton06,DeLucia06,Hopkins06,Lapi06,Shankar06,Monaco07,Somerville08SAM,Cook09,Fontanot09} still do not completely agree on the type of
evolution undergone by massive galaxies (see also \citealt{Dekel09}), on the fraction
of stellar mass formed in the initial, gas-rich burst of star formation,
and on the role played by the late evolution driven
by major and minor mergers.
Nevertheless, all models
agree that galaxies must have been much more compact at the
epoch of formation,
owing to a denser universe, larger gas fractions in the progenitors, and more dissipation.
The latter prediction
has been confirmed by a number of
deep observations of high redshift galaxies \citep[e.g.,][]{Trujillo06,Vandokkum08,Cimatti08,Saracco08},
which have independently found early-type, high-redshift, massive galaxies to be
a factor of a few more compact than local counterparts of the same stellar mass.
Note, however, that several observational biases might
limit the quality and reliability of some of these measurements \citep[e.g.,][]{HopkinsRez,Mancini09,vandokkum09ALL}.

It is still debated how these compact galaxies have
evolved from high-redshifts increasing their
sizes in a way to fall on the size-mass relation we observe today.
As already extensively discussed by, e.g., \citet{Shen03}, \citet{ShankarBernardi}, \PapI,
the scatter around the local, median \re-\mstar\ relation decreases with
increasing mass and, at high stellar masses, is nearly independent of the age
of the galaxies, a challenge for most galaxy evolution models.
In particular, older galaxies are observed to have a steeper size-mass relation than younger systems.

One possible model put forward to explain
the strong size evolution of the red, massive
high-$z$ galaxies is a sequence of minor, dry mergers. Such dynamical
events can ``puff-up'' galaxies by adding mass
in their outskirts, efficiently increasing their sizes, although
limiting the growth of the stellar mass within a factor of $\sim 2$
\citep[e.g.,][]{KochfarSilk06origin,CiottiReview,Bernardi09,Bezanson09,Cimatti09,Hopkins09CompactGalx,Naab09,vanderwel09}.
Recent numerical simulations have however shed doubts on the
actual efficiency of mergers in significantly puffing-up compact
galaxies, and coherently bringing them
along the rather tight structural relations
observed in the local Universe \citep[][]{Nipoti09}.
Moreover, hierarchical models suffer from the serious problem of failing
in fully reproducing the local
size-luminosity relation (see \citealt{GonzalezSAM08}, \PapInp, and references therein).
Another model proposed
in the recent Literature to increase the sizes of massive galaxies is by \citet{Fan08}.
They postulate that the evolution of compact galaxies
undergoes a two-phase
\emph{expansion}: a first one caused by the sudden mass loss via quasar
feedback, and a second one due to the slow mass loss
via stellar winds during which the system slowly evolves towards
a new equilibrium. Their model is broadly consistent with the data
on the local size-mass relation.

Given the large degree of freedom and significant uncertainties in current models of galaxy formation,
it is necessary to look for other ways to test the validity
of a given theory and/or discern ways to improve it. The aim of this paper is
to provide such tests. We will show that the combined
comparison with the size and mass distribution
function of local galaxies can reveal interesting
information on how to improve models of galaxy formation.
In particular, in this work we will use a
detailed hierarchical model of galaxy formation,
show its failures and successes against available data,
and discuss ways to improve it.

We start in \S~\ref{sec|data} describing the data
set we used.
In \S~\ref{sec|models} we describe
the hierarchical model adopted
in this paper, and present its predicted size and mass
distributions for spheroid-dominated galaxies.
We will discuss in some detail the origin
of the discrepancies between model predictions
and the data, and possible ways to improve the model.
We further discuss other aspects of the model in \S~\ref{sec|discu},
and conclude in \S~\ref{sec|conclu}.

\section{DATA}
\label{sec|data}

The sample was extracted from the Data Release 4 of
the Sloan Digital Sky Survey (SDSS; \citealt{York00})
with parameters updated to the Data Release 6.
The sample is magnitude limited to $r$-band deVaucouleur
apparent magnitude $14.5\lesssim m_r \lesssim 17.5$.
Spheroid-dominated objects were selected by requiring
the concentration index to be larger than 2.86 (\citealt{Nakamura03}).
In the following we will take
this dataset as our reference sample for
early-type galaxies to compare the models
with. This sample contains $\sim 70,000$ galaxies and
it extends over a redshift range $0.013 < z < 0.25$, which
corresponds to a maximum lookback time of 3 Gyr.
The effective radii and magnitudes computed by the SDSS
reduction pipeline system were corrected for sky subtraction
problems following \citet{Hyde09a}.
However, as detailed in Bernardi et al. (2009), a sample
selection based solely on concentration,
inevitably contains some S0 and Sa galaxies.
More careful cuts that provide cleaner samples are however possible
\citep{Hyde09a,Hyde09b}. Therefore, in the following
we will also discuss the comparison with the
\citet{Hyde09a} sample dominated by ellipticals.
The comparison with the latter sample is particularly
meaningful for our purposes of understanding the evolution
of the high-$z$, spheroid-dominated, compact, red and dead galaxies
discussed in \S~\ref{sec|intro}.

\section{HIERARCHICAL MODELS OF GALAXY FORMATION}
\label{sec|models}

\subsection{OVERVIEW OF THE MODEL}
\label{subsec|modelsdescription}

Bernardi et al. (2009) showed that the
$V/V_{\rm max}$
method applied to the samples discussed in \S~\ref{sec|data},
yields the filled and open squares shown in
Figure~\ref{fig|Models} for the SDSS
early-type galaxies with concentration\footnote{Defined to be the ratio of the scale which contains 90\% of
the Petrosian light in the $r$-band, to that which contains
50\%.} $C_r>2.86$ (top panels)
and those which satisfy the \citet{Hyde09a} selection
criteria (bottom panels). The left panels
of Figure~\ref{fig|Models} (discussed in \S~\ref{subsec|comparingwiththeERFandSMF}) show the \re\ functions (ERF hereafter), while
the right panels the stellar mass functions (SMF hereafter) of the two samples.
We defer the reader
to Bernardi et al. (2009) for full details on the statistical
and systematic uncertainties in computing a reliable
estimate of the size and stellar mass functions of early-type galaxies.
We just stress here that
the $V/V_{\rm max}$ method is one of the most appropriate
and widely used
techniques to compute the statistical distribution of galaxies
of a given property (either luminosity, size, or velocity dispersion)
among a flux-limited sample (see, e.g., also \citealt{Sheth03}).
For example, as detailed in Appendix~\ref{sec|convolutionofERF},
computing the ERF via some convolution of the luminosity function or the velocity dispersion function,
provides consistent, although not as accurate, descriptions of the ERF.

\begin{figure*}
\includegraphics[width=13truecm]{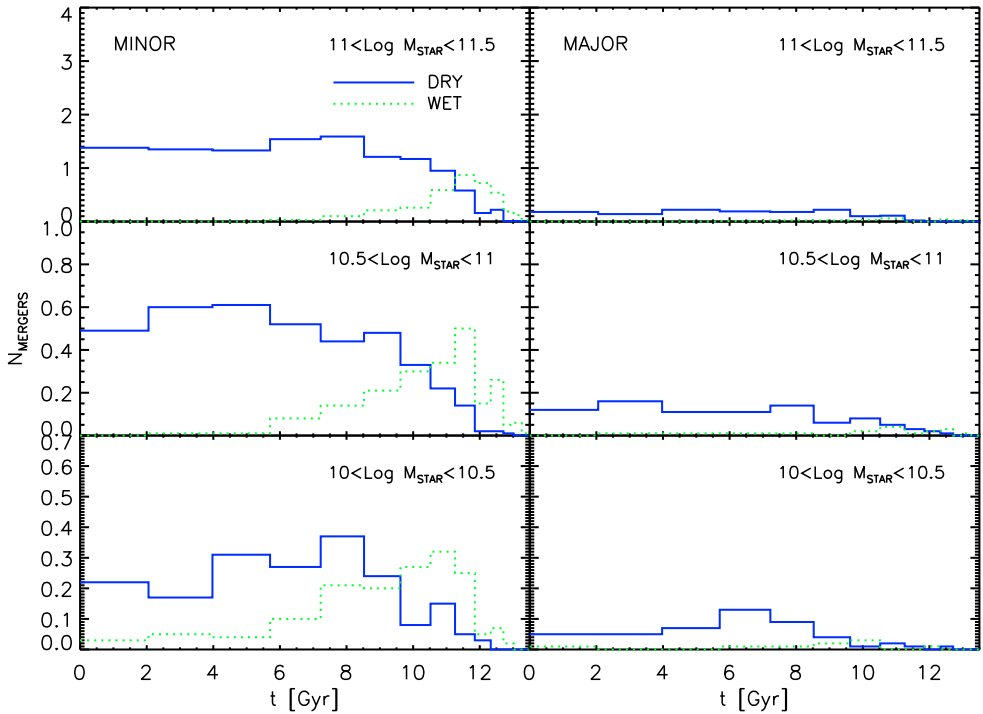}
{\caption{Comparison between the mean number of wet and dry mergers
as a function of lookback time extracted from the merger tress of the Bower et al. (2006) catalog. Each row shows the merger history, averaged
over 100 realizations, of galaxies with stellar mass at $z=0$ in
three different mass bins, as labeled. In the left column we plot
the mean number of minor mergers, with mass ratio $<1:3$, while the
right column shows the mean number of major mergers with mass ratio
$>1:3$. The \emph{dotted} and \emph{solid} lines refer to the mean number of wet and
dry mergers, defined to have a (cold) gas-to-total mass fraction in
the progenitors higher and lower than 0.15,
respectively.}\label{fig|NumberMergersFromModels}}
\end{figure*}

We compare our empirical determination of the ERF with the predictions of the
\citet{Bower06} hierarchical model (the `Durham' model). We will also briefly mention
some characteristics of the \citet[][the `MPA model']{DeLucia06} model, although
we will not compare directly with their predictions as no
available measure of the spheroid sizes are available
from their catalogs at the moment of writing.
These models (see \citealt{Parry08} for a detailed description and
comparison) follow the cosmological co-evolution of
dark matter halos, subhalos, galaxies and supermassive black holes
within the concordance $\Lambda$CDM cosmology. Both models have been
implemented on top of the large, high-resolution cosmological N-body
simulation {\tt MILLENNIUM RUN} (\citealt{Springel05}). The simulation
follows the evolution of $N= 2160^3$ dark matter particles of mass
$8.6\times10^{8}\,h^{-1}{\rm M}_{\odot}$, within a co-moving box of
size $500\, h^{-1}$Mpc on a side and force resolution of 5 \kpch,
from $z=127$ to the present. The cosmological parameters adopted
($\Omega_{\rm m}=0.25$, $\Omega_{\rm b}=0.045$, $h=0.73$,
$\Omega_\Lambda=0.75$, $n=1$, and $\sigma_8=0.9$), are consistent
with the combined analysis of the 2dFGRS (\citealt{Colless01}) and
first year WMAP data (\citealt{Spergel03}). The high mass and
spatial resolution of the Millennium Simulation allows to track the
motion of dark matter substructures inside massive halos, making it
possible to construct merging history trees of all the dark matter
halos and subhalos inside the simulation box (\citealt{Springel05}).
The dynamical evolution of all satellite galaxies is followed until
tidal truncation and stripping disrupt their host dark matter
subhalos. Then a residual survival time is estimated by computing
the dynamical friction formula.

To populate the dark matter subhalos with galaxies and black holes,
both models adopt a set of equations to describe the radiative
cooling of gas, the star formation, metal enrichment and supernovae
feedback, the growth and feedback of supermassive black holes, the
UV background reionization, and the effects of galaxy mergers. A
full description of the MPA and Durham models and a comparison
between their main predictions with observations can be found in
\citet{Bower06}, \citet{Croton06}, \citet{DeLucia06,DeLucia07},
\citet{Malbon07}, \citet{Marulli08}.  In the
following, we will briefly summarize the main physical assumptions
about the formation and evolution of spheroidal galaxies.

In both models the morphology of a galaxy is determined by the
bulge-to-total ratio of its absolute, rest frame luminosity.
Usually, a galaxy is classified as early type if it has $M_{\rm
bulge}-M_{\rm total}<0.4$, where $M_{\rm bulge}$ and $M_{\rm total}$
are the $B$-band magnitude of the bulge and of the whole galaxy,
respectively.  The star formation is assumed to be proportional to
the galaxy cold gas mass.  Stellar winds and supernovae reheat a gas
mass proportional to the mass of the stars, increase the metallicity
of the disk material and inject substantial amounts of energy into
their surroundings.  Both models use the stellar population
synthesis model of \citet{BruzualCharlot03} to describe the stellar
population properties.

The evolution of early type galaxies is then regulated by galaxy
mergers, disk instabilities and AGN feedback.  It is assumed that
galaxy {\em major} mergers ($M_{\rm gal,2}/M_{\rm gal,1}\ge 0.3$)
disrupt any stellar disk present and produce a spheroidal remnant,
which contain all the stars of its progenitor galaxies. The gas
present in the merging galaxies forms stars in a burst, which are
then added to the new spheroid. In the case of a {\em minor} merger
($M_{\rm gal,2}/M_{\rm gal,1}<0.3$), the disk of the primary galaxy
survives, and the stars and the gas of the satellite are added to
the bulge and to the disk of the primary galaxy, respectively. While
in the MPA model starbursts occur in all minor mergers, in the
Durham model there is a burst only if $M_{\rm gal,2}/M_{\rm
gal,1}\ge0.1$ and if the primary galaxy has $M_{\rm gas}/M_{\rm
disk}>0.1$. In both models AGN feedback in the radio mode further
reduces or even stops late cooling flows in the halo centers.  This
behavior plays a key role in reproducing the exponential cut-off in
the bright end of the local galaxy luminosity function, and in
determining the bulge-dominated morphologies and old stellar ages of
the most massive galaxies in clusters.

One key ingredient of hierarchical galaxy formation models of
relevance for this paper, is how massive spheroids grow in size after
they form. We here recap some of the basic model assumptions and
refer to, e.g., \citet{GonzalezSAM08} for further details. The
radius of the merger remnant is computed from energy conservation.
Assuming virial equilibrium, the radius of the remnant is estimated
solving the equation $E_{\rm after}=E_1+E_2+E_{\rm orb}$, with $E_1$
and $E_2$ the total internal energies of the two merging
progenitors, and $E_{\rm orb}$ their orbital energy. Each galaxy is
characterized by its mass and radius, estimated assuming a $R^{1/n}$
profile, with $n=4$ usually assumed for early-type galaxies. The
energy of each galaxy is simply written as $E_i=kGM_i^2/R_i$, with
$M_i$ the \emph{total} mass within its radius, $k$ a constant
depending on the profile, and $R_i$ the half-mass radius.
From the virial condition, the radius of the remnant obeys the condition
\begin{equation}
R_f = \frac{M_f^2}{\frac{M_1^2}{R_1}+\frac{M_2^2}{R_2}+2f_{\rm orb} \frac{M_1 M_2}{R_1+R_2}}\, ,
\label{eq|virialcondition}
\end{equation}
with $0 \lesssim f_{\rm orb}\lesssim 2$ parameterizing the (uncertain) orbital energy of the progenitors \citep[e.g.,][]{Cole00}.

\subsection{THE GROWTH OF EARLY-TYPE GALAXIES IN THE MODEL}
\label{subsec|mergers}

\begin{figure*}
\includegraphics[width=13truecm]{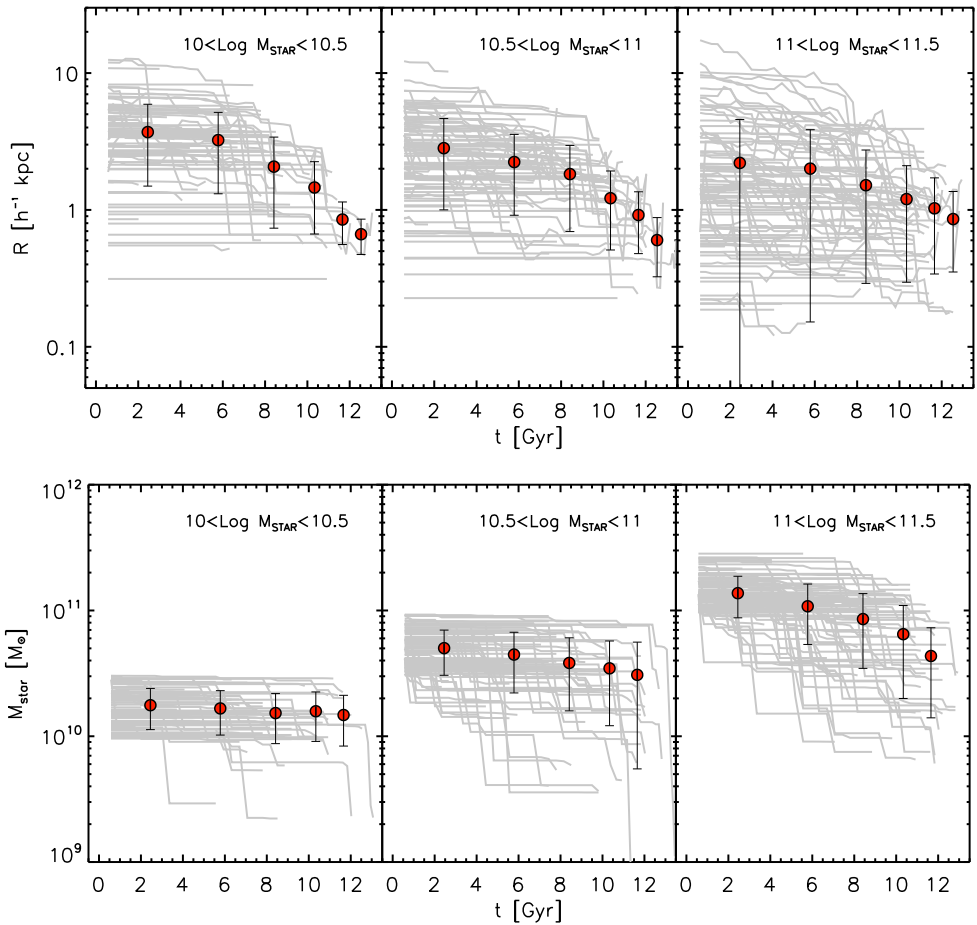}
{\caption{Hundred merger histories for the
evolution of the size (upper panels) and stellar mass (lower panel)
as a function of time for galaxies of
different mass at $z=0$ as predicted by the Bower et al. (2006)
model. The median and variances of the merger histories are shown in
each panel with \emph{red circles} with error bars.
}\label{fig|RezModels}}
\end{figure*}

One of the main drivers for the strong evolution of bulge-dominated galaxies in size and
stellar mass in hierarchical models is mergers (see also discussions in, e.g.,
\citealt{DeLucia06,DeLucia07,Parry08}).
Figure~\ref{fig|NumberMergersFromModels} shows
the mean number $N_{\rm MERGERS}$ of
wet and dry mergers per Gyr a galaxy had since its
formation epoch as a function of lookback time $t$.
Each row shows the mean merger history, averaged over 100
realizations of the merger trees in the \citet{Bower06} catalog,
of galaxies with stellar mass at $z=0$ residing in the
mass bin indicated at the top of each panel, as labeled. In the left
column we plot the mean number of minor mergers, with mass ratio
$<1:3$, while the right column shows the mean number of major
mergers with mass ratio $>1:3$. The dotted and solid lines refer to
the mean number of wet and dry mergers defined to have a cold
gas-to-total mass fraction in the progenitors higher and lower than
0.15, respectively.
On average, the number of dry mergers grows with
final stellar mass, and galaxies that end up with
\mstar$> 10^{11}\, $\msun\, tend to undergo significantly
more numerous merging events than galaxies with lower final
stellar mass.

We can sketch a general trend for the evolution
of massive early-type galaxies in hierarchical
models. A large portion of early-type galaxies is usually formed at high redshifts
through a wet merger of gas-rich disk progenitors and then continued
accreting stellar mass mainly through minor, dry mergers. Therefore,
the epoch of formation is generally identified to be at high-$z$,
when the first starburst event took place and the central potential
well was settled. We also find that a significant
fraction of massive bulges
is also formed through instability of gas-rich
disk galaxies. Even in this case the initial
size of the newly formed bulge is set by virial equilibrium
and energy conservation, adopting a condition
very similar to Eq.~\ref{eq|virialcondition}
(see \citealt{Cole00} for details). Irrespective of their
exact formation process, most of the massive spheroids in the model kept on
accreting mass and increasing their sizes through dry mergers,
largely extending the epoch of stellar mass assembly. The \citet{Bower06} model
predicts that all massive early-type galaxies, on average, undergo $\sim 3-7$
minor mergers and $\lesssim 1$ major dry mergers since their
formation epoch (i.e., identified as the epoch of the major wet merger
among the gas-rich disk progenitors).

These late evolutionary
features are a general trend for most
hierarchical models. For example, we have checked that
the \citet{DeLucia06} model (see also \citealt{GuoWhite08}) predicts
a similar pattern for the growth of early-type galaxies,
first characterized by a wet, formation phase peaked at high-redshifts, and a subsequent
evolution dominated by a series of minor, dry mergers. The latter model
similarly also predicts a rate of mergers increasing with
increasing final stellar mass.

The grey lines in the upper and lower panels of Figure~\ref{fig|RezModels} show how the sizes and masses, respectively, change
for a hundred merger histories drawn from the \citet{Bower06} catalog, so that
the final stellar mass at $z=0$ will fall within the mass bin labeled
at the top of each panel.  In each case the trees are followed back
in time, choosing the most massive early-type progenitor, until
this is no longer possible. We have checked that our randomly selected
bulge-dominated galaxies in the model have mainly grown in mass through mergers,
with disk instabilities increasing the spheroid masses by only $\lesssim 10\%$, decreasing to
$\lesssim 2\%$ for the most massive galaxies.
This is clearly visible from the lower panels of Figure~\ref{fig|RezModels},
where most of the stellar mass accretion histories remain almost flat most
of the time and have sudden jumps in correspondence of merging events.

Figure~\ref{fig|RezModels} shows that only a small fraction
of low mass objects (with \mstar$\, < 10^{11}\,$\msun) were early-types 10~Gyrs ago, but their mass
has changed little since they first became early types ($\lesssim 50$\%).  The sizes
of this population of galaxies, however, have grown by at least a
factor of three over this time, as expected
if their evolution is driven by minor mergers.
In contrast, a larger fraction of massive galaxies had formed 10~Gyrs ago; however,
whereas the mass of the population has grown
by about a factor of three since then (with most of the growth
occurring at lookback times greater than about 6~Gyrs), the sizes
have grown by a smaller factor ($\sim 2$). Indeed, the models predict that
a substantial fraction of massive galaxies (with \mstar$\, \gtrsim 10^{11}\,$\msun)  have not changed in size
for a long time. More specifically, in the highest mass bin considered,
we find that about 21\% of the galaxies
have efficiently increased their sizes by a factor greater than 3,
a significant fraction of 22\%
remains instead extremely compact (although still growing in mass),
and the rest undergoes
a milder evolution
with a size increase contained within a factor $< 3$.
We have verified that while most ($\sim 60\%$) of the mass
growth is added via major mergers,
most ($\sim 90\%$) of the size growth, where this actually happens,
is via \emph{minor} mergers. As extensively discussed
in \S~\ref{subsec|comparingwiththeERFandSMF}, the survival of a non-negligible fraction of
compact and massive galaxies until the present epoch is at variance
with the SDSS distribution of early-type galaxies (with B/T $\gtrsim 0.5$),
and we will discuss possible causes and improvements to this problem in \S~\ref{subsec|progenitors}.
However, we will also
show that when restricting the analysis to spheroid-dominated systems (with B/T$>0.7$),
the predicted number density of ellipticals with size $\sim 1$ kpc
is \emph{consistent} with the data.

\subsection{COMPARISON WITH THE SIZE AND STELLAR MASS DISTRIBUTIONS}
\label{subsec|comparingwiththeERFandSMF}

In this section we compare the data on the size and mass distributions of SDSS early-type
galaxies with the predictions
of SAMs. As partly explained above, it is beyond the scope of the
current paper to build
an ab initio complete hierarchical model able to predict the sizes, stellar masses
and velocity dispersions of early-type galaxies, which
we postpone to future work.
Instead, the aim here is to
show the what we can learn from the simultaneous comparison with the size and
mass distribution of early-type galaxies.

\subsubsection{THE PREDICTED SIZE-MASS RELATION}
\label{subsubsec|ReMstarRelation}

The orange, green, and blue contour
levels in Figure~\ref{fig|SizeMassRelation} indicate the region of the size-mass
plane containing 68\%, 95\%, and 99.7\% of the whole SDSS sample of local
early-type galaxies, respectively.
All the contour levels have been weighted by the appropriate $1/V_{\rm max}$, as in \citet{Bernardi09} and \citet{ShankarBernardi}. The cyan, yellow, and red contour levels contain the 99.7\%, 95\% and 68\% of the
corresponding mock sample of local early-type galaxies from \citet{Bower06}. Figure~\ref{fig|SizeMassRelation}
plots the \citet{Hyde09a} sample, and therefore, for consistency,
we only adopt mock galaxies with $B/T>0.7$. We have also
checked that the mean trend and scatter in the predicted size-mass relation
does not change significantly when adopting lower limits
of the bulge component, such as $B/T>0.5$. The solid squares
with error bars show the predicted median and variances of the sizes at fixed stellar mass for the models. We find that the model predicts an increasing size
when moving from the lower to the highest masses
up to \mstar$\sim 2\times 10^9$\msun.
However, as seen in the Figure, at stellar masses above
\mstar$\sim 2\times 10^9$\msun, at variance with the data
the model predicts a strong
flattening in the predicted size-mass relation,
while at \mstar$\sim 3 \times 10^{11}$\msun, the sizes
start increasing again with increasing stellar mass.
Similar findings were also recently discussed
by \citet{GonzalezSAM08}.

Not only the observed and predicted size-mass distributions differ
in slope and zero point, but also the predicted
scatter (in size at fixed stellar mass or vice versa), is much larger than the observed one.
Figure~\ref{fig|SizeMassRelation} clearly highlights the problem in the predicted size-mass relation, already noticed by some previous
studies (e.g., \citealt{GonzalezSAM08}, \PapInp).
We will show below that
despite some possible improvements towards reproducing
the size and mass functions, the
full match to the size-mass
relation remains an extremely non-trivial task for the model.

\subsubsection{THE PREDICTED SIZE FUNCTION}
\label{subsubsec|matchERF}

\begin{figure}
\includegraphics[width=8.5truecm]{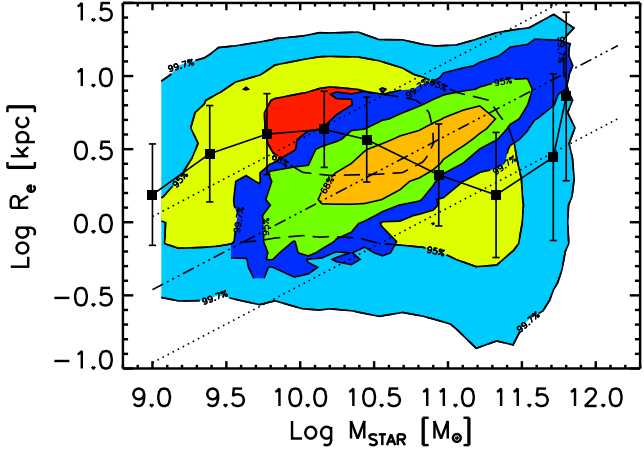}
\caption{Predicted size-mass relation for the \citet{Bower06} sample, with the red, yellow, and blue contours marking the
region of the \re-\mstar\ plane containing 68\%, 95\%, and 99.7\% of the mock catalog, respectively, and the solid
squares indicating the median \re\ at fixed stellar mass and its standard deviation.
Also shown the observed size-mass relation for the SDSS sample with the orange, green, and blue contours marking the
region of plane containing 68\%, 95\%, and 99.7\% fraction of the total sample.
The predicted size mass relation flattens at high-masses, at variance with the data, although still
a significant comoving number density of large galaxies is predicted by the model.
The dot-dashed line
in the figure is the best-fit relation from \PapInp, while the dotted lines
roughly mark the locus of points 3-$\sigma$ away from the median relation.}
\label{fig|SizeMassRelation}
\end{figure}

\begin{figure*}
\includegraphics[width=17.5truecm]{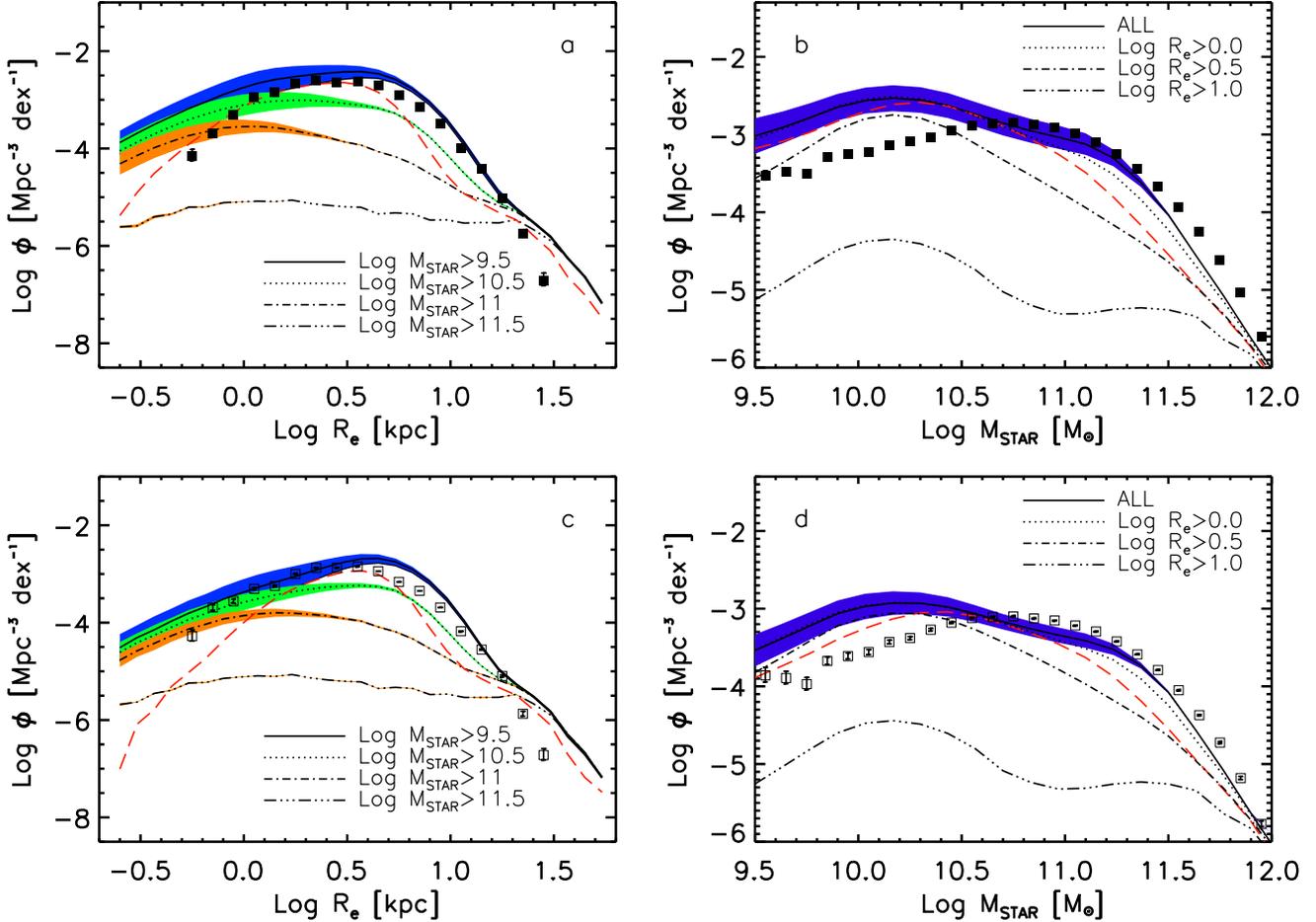}
\caption{\emph{Upper left panel}: comparison of the measured ERF for galaxies
with $C_r>2.86$ (\emph{filled squares}) from Bernardi et al. (2009), compared to
the one predicted by the \citet{Bower06} model.
The different lines
show the contribution of galaxies from the model with masses above a different threshold, as labeled, and with $B/T>0.5$. The colored areas around each curve mark the uncertainty in defining
an early-type galaxy in the model with the
color regions bracketing the galaxies with
$B/T>0.3$ and $B/T>0.7$. The red \emph{long-dashed} line marks the predictions of a model in which the galaxies with sizes
below and above
the \emph{dotted} lines in Figure~\ref{fig|SizeMassRelation} are removed from the mock catalog. 
\emph{Lower left panel}: same format as panel \emph{a}
but now restring to the ERF computed from the \citet{Hyde09a} sample (\emph{open squares}), dominated by ellipticals, compared to the predicted ERF for galaxies with $B/T>0.75$.
\emph{Right upper panel}: comparison between the measured and predicted stellar mass functions for samples as in panel \emph{a}. The data (\emph{filled squares}) refer to the
sample of galaxies with $C_r>2.86$. The \emph{solid} line identifies the contribution of all early-type galaxies in the model with $B/T>0.5$, while the red \emph{long-dashed} line considers
only $B/T>0.5$ galaxies within the \emph{dotted} lines in Figure~\ref{fig|SizeMassRelation}. The other lines show the contribution of galaxies with effective radius above a different threshold, as labeled. \emph{Right lower panel}: same format as panel \emph{b} but
referred to the stellar mass function from the \citet{Hyde09a} sample (\emph{open squares}), and for model galaxies with $B/T>0.75$. The colored areas around each curve mark the uncertainty in defining
an early-type galaxy in the model with the
color regions bracketing the galaxies with
$B/T>0.55$ and $B/T>0.9$. It is clear from these plots that simply
removing the compact galaxies from the sample severely underpredicts the high-mass
end of the stellar mass function.}
\label{fig|Models}
\end{figure*}

The blue area in Figure~\ref{fig|Models}\emph{a} shows the \citet{Bower06}
predicted size distribution for early-type galaxies computed by counting the number of sources within a given bin of size $R$ divided by the volume of the Millennium simulation, (500$\, h^{-1}\,$ Mpc on a side, where we set $h=0.7$ to get the number density in units of Mpc$^3$). Following \citet{GonzalezSAM08}, we further correct the half-mass radius $R$ by a factor of 1.35, assuming that all galaxies strictly follow a de Vaucouleurs profile (see \citealt{GonzalezSAM08} and references therein).
We further divide the sizes by a factor of 0.7 to scale from the units of $h^{-1}$ kpc in the model to kpc in the data. We also divide the masses in units of $h^{-1}$\msun\ by 0.7 and increased them by 0.05 dex to account for the slightly different initial mass functions adopted in the model (Kennicutt 1983) and the data (Chabrier 2003; see Table 2 in Bernardi et al. 2009). We then select from the \citet{Bower06} catalog all those galaxies with a total stellar
mass above $\log M_{\rm star}/M_{\odot}\sim 9.5$, consistent with our sample, and a prominent bulge component.
The blue area in Figure~\ref{fig|Models}\emph{a} includes all galaxies with $M_{B, {\, \rm bulge}}-M_{B,{\, \rm tot}}<0.4$ (lower limit) and $M_{B, {\, \rm bulge}}-M_{B,{\, \rm tot}}<1.3$ (upper limit), which corresponds to the subsample of galaxies with a bulge-to-total stellar mass ratio of
$B/T= 0.5\pm 0.2$ (see, e.g., \citealt{Cole00}, \citealt{Laurikainen09} and references therein). As extensively discussed in \citet{GonzalezSAM08},
a cut in the mock catalog of $B/T \le 0.5$ is equivalent to a cut in concentration of $C_r>2.86$, the
one adopted by Bernardi et al. (2009) to classify early-type galaxies in the sample
shown with solid, filled squares in Figures~\ref{fig|Models}\emph{a},\emph{b}.

Overall, the model predicts a nearly Gaussian-shaped size distribution,
with a peak around $\sim 3$ kpc, which quite closely resembles
the observed one. In particular, the model
predicts a sufficiently large number of intermediate-size
and large galaxies (\re$\gtrsim$ 1 kpc)
in broad agreement with data.
Such a result is not
trivial. The outputs of hierarchical models concerning the sizes
of galaxies have been discussed several times in the Literature (see \S~\ref{sec|intro}).
However, emphasis has been usually put on the tendency of the models
to predict too many compact (\re $\lesssim 1 $ kpc) galaxies, but not much
was discussed about larger galaxies.
We confirm indeed that, with respect to the data, the hierarchical model considered here
(see solid line in Figure~\ref{fig|Models}\emph{a}), shares this common feature of
predicting too many, up to a factor of $\sim 10$ higher, early-type galaxies with sizes below $\sim 1$ kpc
and stellar mass \mstar$\gtrsim 3\times 10^{10}$\msun.
On the other hand, it
also predicts the number density of larger galaxies
to be somewhat in better agreement with SDSS measurements, although
it still overpredicts by a factor $\sim 2$ the number density of
galaxies with sizes between \re$\sim 3-5$ kpc, and even more that of
ultra-large galaxies with \re$\gtrsim 15$ kpc.
These outcomes of the model would obviously not have been evident a priori by just
plotting the size-mass relation, which predicts a rather flat distribution
in the median sizes, and it lends support on the importance
of comparing model outputs with size distributions as well as mass functions.

Bernardi et al. (2009) showed
that the $C_r>2.86$ sample has a significant
contribution from S0 galaxies ($\gtrsim 30\%$). The latter class of galaxies has light profiles
best-fitted by a combination of an exponential and a $r^{-1/4}$ de Vaucouleurs' profile (e.g., \citealt{deVac}), which
yields somewhat larger effective radii (by $\sim 0.1$ dex) than a pure $r^{-1/4}$ profile.
This in turn might bias the result of Figure~\ref{fig|Models}\emph{a}, because by assuming
a pure $r^{-1/4}$ profile for all early-type galaxies, we could have
overpredicted the number of compact galaxies. We have however
checked that even after
increasing the sizes of all S0 mock galaxies (those with $B/T \lesssim 0.7$) by 0.1 dex to match
the results of Bernardi et al. (2009), the predicted ERF is very similar.
We therefore conclude that the overproduction of early-type galaxies
at small scales is real and inherent in the model itself.

\begin{figure*}
\includegraphics[width=17.5truecm]{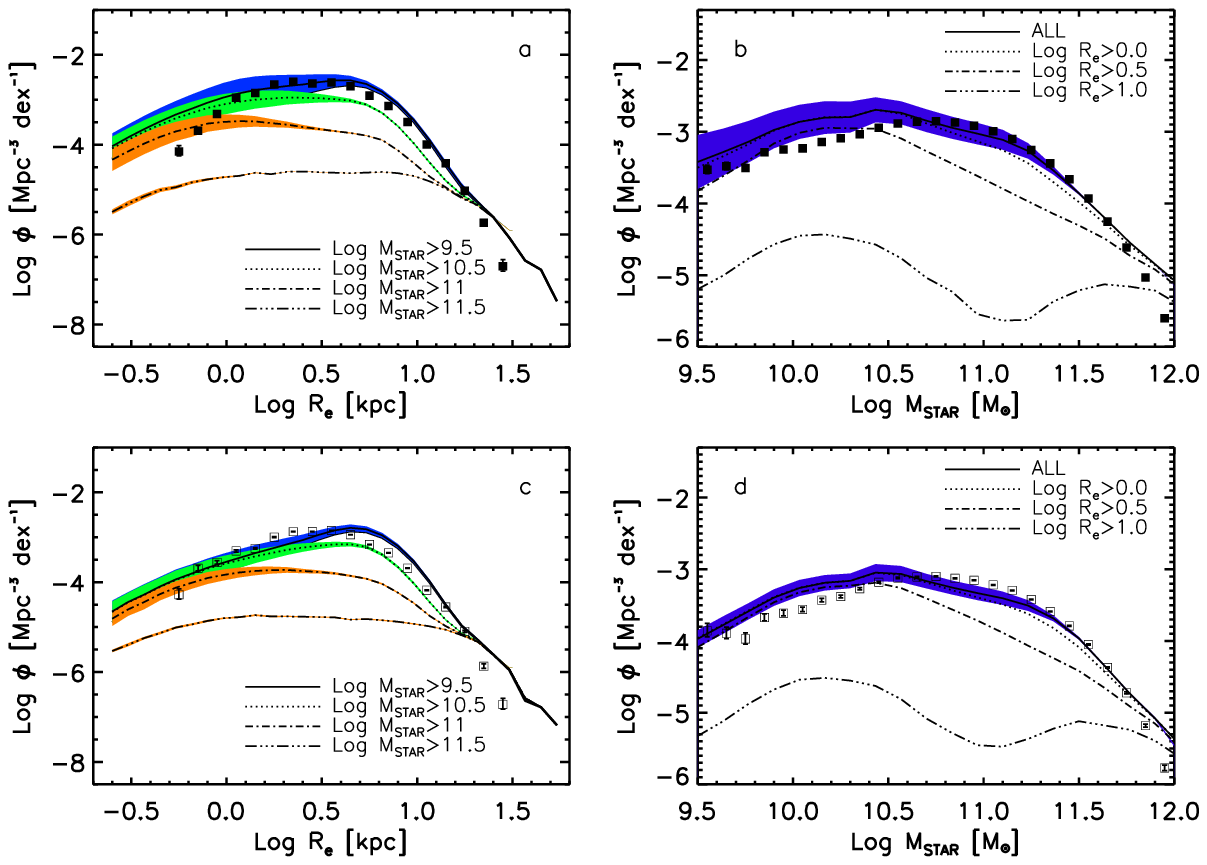}
\caption{Same format as Figure~\ref{fig|Models}, but here
the model predictions have been modified by assuming
that satellite galaxies with stellar mass below \mstar$\sim 3\times 10^{10}$\msun\
have merged with the central galaxy. It is clear that increasing
the rate of mergers of the less massive spheroids yields a considerably
better agreement with the data, although it does not improve
the match to the size-mass relation (see Figure~\ref{fig|SizeMassRelationPlus}).}
\label{fig|ModelsImproved}
\end{figure*}

Interestingly, Figure~\ref{fig|Models}\emph{c} shows that
when comparing to the elliptical-dominated sample by \citet{Hyde09a},
characterized by a minimal contribution from S0 and Sa galaxies (open squares),
the overproduction of compact galaxies is significantly reduced, and possibly
confined to only ultracompact (\re$\lesssim 0.5$ kpc) galaxies.
For this comparison we used mock galaxies with $B/T>0.75$, roughly the minimum
$B/T$ in the \citet{Hyde09a} sample (solid line with blue area).
When moving from the $C_r>2.86$ to the \citet{Hyde09a} sample,
the shapes of the observed and predicted ERF are similar:
while the observed ERF only slightly decreases in normalization,
the predicted size function significantly drops below $\lesssim 1$ kpc,
better matching the data.

\citet{Trujillo09} (see also \citealt{Taylor09}) have recently
claimed that in the SDSS Data Release 6, only
a tiny fraction ($\sim 0.03\%$) of compact ($\lesssim 1.5$ kpc)
and massive (\mstar$\gtrsim 8\times 10^{10}\, $\msun) systems is
present in the local Universe ($z<0.2$). \citet{Trujillo09} also
claim that this estimate is actually an upper limit,
given that most of these galaxies are
metal-rich and relatively young ($\sim 2$ Gyr). They therefore conclude
that the nearly complete absence of very compact and old galaxies in the nearby
Universe, is at variance with predictions of detailed
hierarchical models that grow galaxies through mergers. The result discussed
in Figure~\ref{fig|Models}\emph{a},\emph{c} are in broad agreement
with the \citet{Trujillo09} findings, that
hierarchical models, or at least the one considered here,
tend to produce too many compact galaxies.
However, we find that such a discrepancy is not
as strong as claimed before, and we also
add that the difference is morphology-dependent, i.e., significantly
reducing with increasing B/T.

On the other hand, when moving to more spheroid-dominated
galaxies, the model
tends to produce an higher fraction of galaxies with \re$\gtrsim 3$ kpc,
predicting up to a factor of $\sim 10$ higher number density of super-large galaxies
(with \re$\gtrsim 30$ kpc). This overproduction of large galaxies with
mass around $\sim 10^{10}$\msun\ is one of the major causes, together with the
overproduction of compact and massive galaxies, for the flattening
of the size-mass relation seen in Figure~\ref{fig|SizeMassRelation}.

\subsubsection{THE PREDICTED STELLAR MASS FUNCTION}
\label{subsubsec|matchSMF}

Figures~\ref{fig|Models}\emph{b,d} show the corresponding comparison
between model predictions and SDSS data on the SMF
for the $C_r$ sample (panel \emph{b}) and the \citet{Hyde09a} sample (panel \emph{d}).
It is apparent that the model
provides a poorer match to the SMF, irrespective of the
sample considered. More noticeably, at variance with the data
the model predicts a ``bump'' in the number density
of early-type galaxies around \mstar$\sim 1.5\times 10^{10}$\msun,
and falls short by a factor of $\sim 2$ in producing enough
galaxies with mass \mstar$\gtrsim 10^{11}$\msun. We have checked that
the model by \citet{DeLucia07}, possibly due to the different treatment of dynamical friction timescales
(see, e.g., \citealt{Parry08,Seek09}), indeed produces a flatter
distribution in the number density of early-type galaxies, in better agreement with
the data.

\subsection{LOOKING INTO THE MODEL}
\label{sec|improve}

\subsubsection{INCREASING THE SATELLITE-CENTRAL MERGER RATE}
\label{subsec|progenitors}

A very preliminary test to improve on the flatness of the
predicted size-mass relation would be to simply remove
from the mock catalog the large and less massive galaxies (\re$\gtrsim 3$ kpc and \mstar$\sim 10^{10}$\msun),
and the compact and massive ones (\re$\lesssim 1$ kpc and \mstar$\sim 10^{11}$\msun).
One simple way
to do this would be to assume that those
galaxies which are strong outliers with respect to the
observed local size-mass relation in Figure~\ref{fig|SizeMassRelation}, should be considered
as irregulars and/or disk-dominated galaxies instead of spheroids (e.g., \citealt{Hopkins09transf}, and references therein).
The region in the \re-\mstar\ plane within the dotted lines
in Figure~\ref{fig|SizeMassRelation} roughly defines the ``region of acceptance'' at a $\sim 3 \sigma$ level. All the mock
galaxies which lie below and above those lines can be safely considered
as outliers. Recomputing the predicted ERF and SMF only considering
the subsample of galaxies within the dotted lines,
yields the red long-dashed lines in Figures~\ref{fig|Models}\emph{a,b,c,d}.
It is clear that, irrespective
of the actual physical basis motivating such drastic cut in the mock catalog of early-type galaxies,
this is not a satisfactory solution, as it severely further
underpredicts the number density of massive galaxies, without
significantly reducing the bump around \mstar$\sim 1.5\times 10^{10}$\msun,
and also worsens the match to the ERF.

We have checked that a significant fraction of the outlier
galaxies are satellites. Therefore,
one possible way to remove the outlier galaxies might
be to simply assume
a higher rate of mergers with the central galaxy in the same parent halo.
Figure ~\ref{fig|ModelsImproved} shows the model predictions
modified by assuming that all
satellite galaxies with stellar mass below \mstar$=3\times 10^{10}$\msun\
have merged with their respective centrals. It is apparent how this hypothetical increase
in the merger rate of the less massive spheroids yields a considerably
better agreement with the observed SMF, lowering the number density
at low masses, and increasing it at higher masses.

\begin{figure}
\includegraphics[width=8.5truecm]{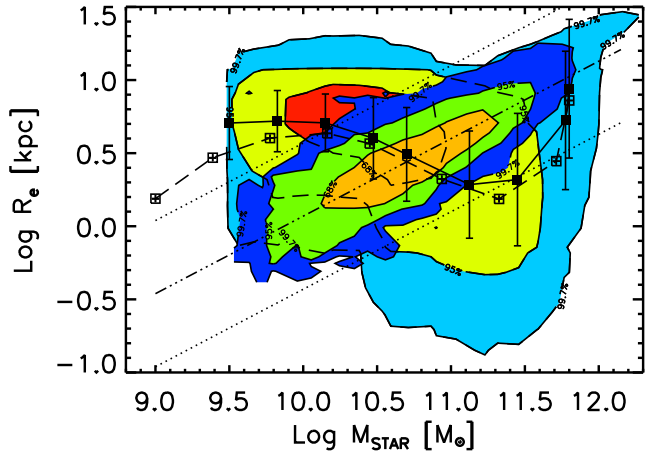}
\caption{Same format as Figure~\ref{fig|SizeMassRelation} with SDSS data compared
to the predicted size-mass relation for the \citet{Bower06} model
with low-mass, type 1 galaxies merged with their centrals, as discussed
in \S~\ref{sec|improve}. For comparison we also
show with a \emph{long-dashed} line the relation from
Figure~\ref{fig|SizeMassRelation}.
The size-mass relation is still not properly reproduced.}
\label{fig|SizeMassRelationPlus}
\end{figure}

Although the simple improvement proposed here for this model
yields outputs in better agreement with the data,
it requires a valid physical explanation
and several independent tests (such as clustering).
Some hints that may motivate a revision of the merger rates
in the model have already been
discussed in the Literature. First of all, there is the observational fact that the frequently assumed
minor merger rate of massive galaxies seems to be lower than what suggested by recent observations (Jogee et al. 2009). Second, although the galaxy merger rates
broadly follow the high-resolution merger rates of the dark matter halos and subhalos in the Millennium simulation, the actual computation of the former may still not be accurate enough. In fact, baryons can affect their surrounding
halos because of, e.g., adiabatic contraction (e.g., \citealt{Tissera09}, and references therein), leading to
a denser subhalo, which would not be as easily tidally stripped, hence allowing for a faster merger. More recently,
several groups working with the same model considered here, have
shown that an increased galaxy merger rate produces better fits to the
luminosity function for different Hubble types (\citealt{Benson09}) and for
the clustering (\citealt{Seek09}).

When  merging the satellite galaxies, we have also
updated the sizes of the centrals following
Eq.~(\ref{eq|virialcondition}) and setting $f_{\rm orb}=2$.
However, as shown
in Figure~\ref{fig|ModelsImproved}\emph{a,c}, at variance with
the SMF, the comparison to the observed ERF does
not significantly improve. Note that using lower values
of $f_{\rm orb}$ would yield
larger sizes for the remnants, thus further worsening the match to the data.
The model continues to predict too many
large and ultra-large galaxies, with respect to those observed.
More importantly, we find that the resulting size-mass relation
for this kind of model still presents similar discrepancies
with respect to the data, as shown in Figure~\ref{fig|SizeMassRelationPlus}.
Low-mass galaxies still have too large sizes, while a non-negligible fraction
of the massive galaxies
still have too compact sizes, thus preserving the flattening of the relation.

\subsubsection{THE SIZE-MASS RELATION AT HIGHER REDSHIFTS}
\label{subsec|remassz}

\begin{figure*}
\includegraphics[width=17.5truecm]{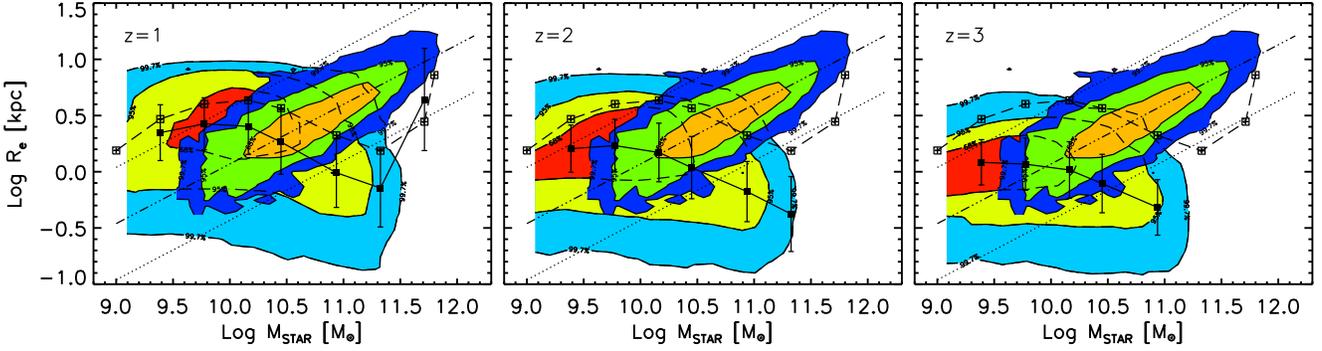}
\caption{Same format as Figure~\ref{fig|SizeMassRelation}
where we plot the predicted size-mass relation of
early-type galaxies (B/T $\gtrsim 0.7$) for the \citet{Bower06} model
at different redshifts, as labeled, and compared
with SDSS data at $z=0$ for reference. The \emph{long-dashed} line
in each panel is the size-mass relation at $z=0$, from Figure~\ref{fig|SizeMassRelation}.
The presence of a flattened
relation at all epochs suggests that some of the problems
might be linked to the initial conditions.}
\label{fig|SizeMassRelationReds}
\end{figure*}

Figure~\ref{fig|SizeMassRelationReds} shows the size-mass relation for ellipticals (B/T $\gtrsim 0.75$)
at different redshifts, as labeled. It is evident
that the flattening discussed above is present at all epochs
(see also Figure 5 in \PapI).
At fixed stellar mass, galaxies with \mstar$< 10^{11}$\msun,
tend to shrink by a factor of $\sim 3$ at higher redshifts. At higher
masses, although a large fraction manages to grow (at $z\lesssim 2$)
in size by a significant
amount enough to saturate (and actually overproduce) the number counts
of large galaxies in the local Universe, still a large portion of it
remains compact, as we further discuss in the following section.

We here note that, with respect to observations, low mass galaxies tend to be already quite
large (\re$\gtrsim 1$ kpc) close to their formation epoch.
As reviewed by \citet{Mancini09},
most of the low-mass galaxies observed
so far in deep surveys, have in fact on average a factor of $\lesssim 2$
smaller sizes than the ones predicted here at the same
redshifts.
This is due a combination of several processes. Some
of the spheroids are generated by strong disk instabilities
in the progenitor. The exact treatment of instabilities
and of the energy conservation between the progenitor disk
and the final bulge, have been discussed
by \citet{Cole00} and \citet{Bower06}, and are
still subject to significant uncertainties that might
require some closer comparison with hydro-simulations to provide
more reliable answers. The spheroids
formed instead through the merging of disk dominated, gas-rich progenitors,
could possibly be more compact than currently assumed in the present
model, as more gas in the progenitors effectively produces more compact
sizes than the ones estimated from dissipationless virial relations
of the type given in
Eq.~\ref{eq|virialcondition} (e.g., \citealt{HopkinsRez}, and references therein).
The second problem of the formation and survival
of compact and massive galaxies is a problem of possibly
different nature and we discuss it in some detail below.

\subsubsection{COMPACT AND MASSIVE GALAXIES}
\label{subsec|progenitors}

As discussed above,
Figures~\ref{fig|Models}\emph{a,b}
reveal that the model tends to predict a
larger number of compact ($\lesssim 1$ kpc) and massive
(\mstar$\gtrsim (0.5-1)\times 10^{11}$\msun) early-type galaxies,
with respect to the data. Such a result persists even after allowing
for an increased rate of mergers for central-satellite
galaxies required to improve the match to the observed SMF (Figure~\ref{fig|ModelsImproved}\emph{a,b}).
The extreme compactness of massive galaxies
plays a considerable role in the flattening of the predicted
size-mass relation shown in Figure~\ref{fig|SizeMassRelation}.
We here attempt to explore possible reasons
responsible for producing smaller sizes, or suppressing their later growth,
in massive galaxies.

\begin{figure*}
\includegraphics[width=7.5truecm]{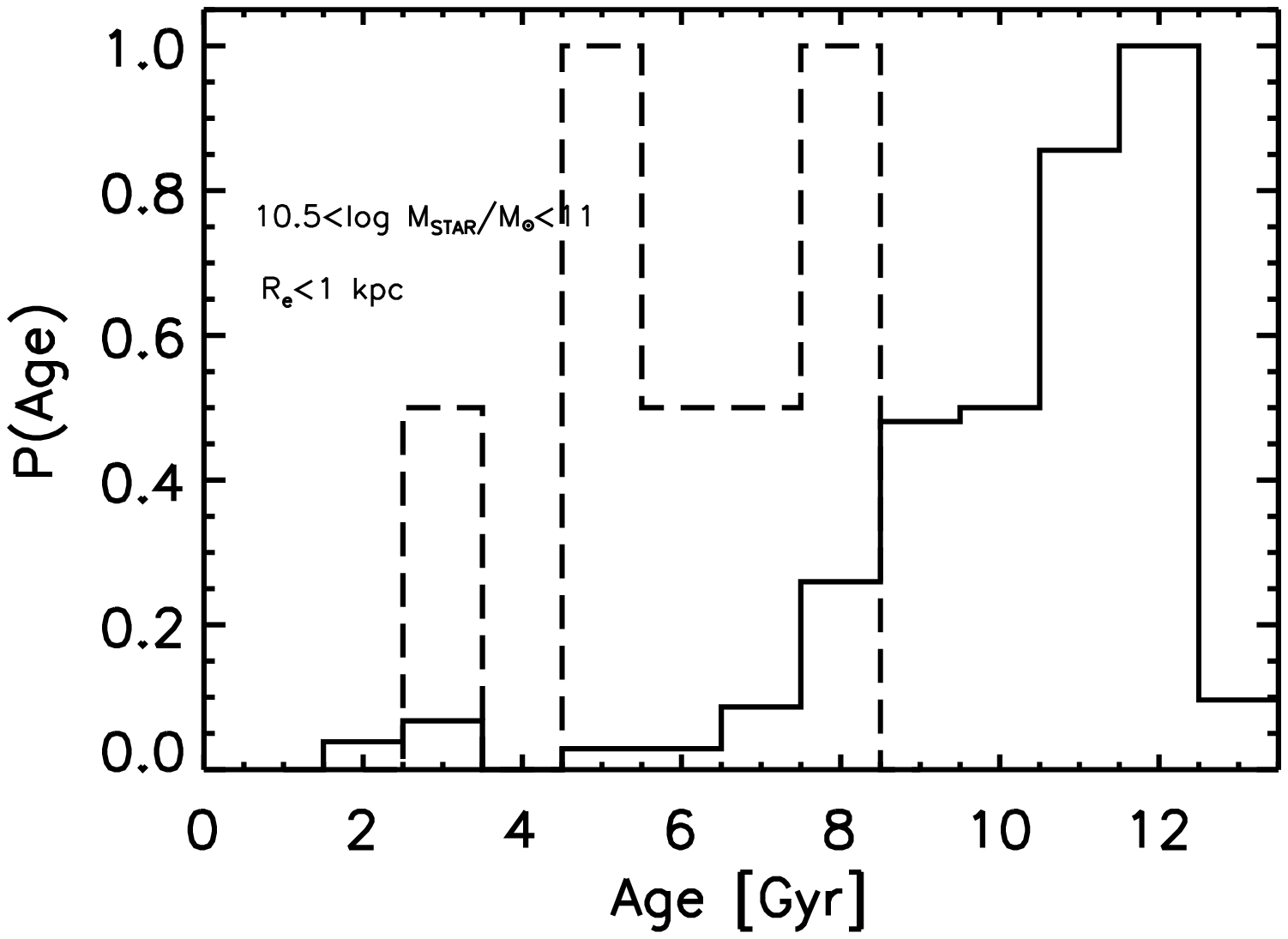}
\includegraphics[width=7.5truecm]{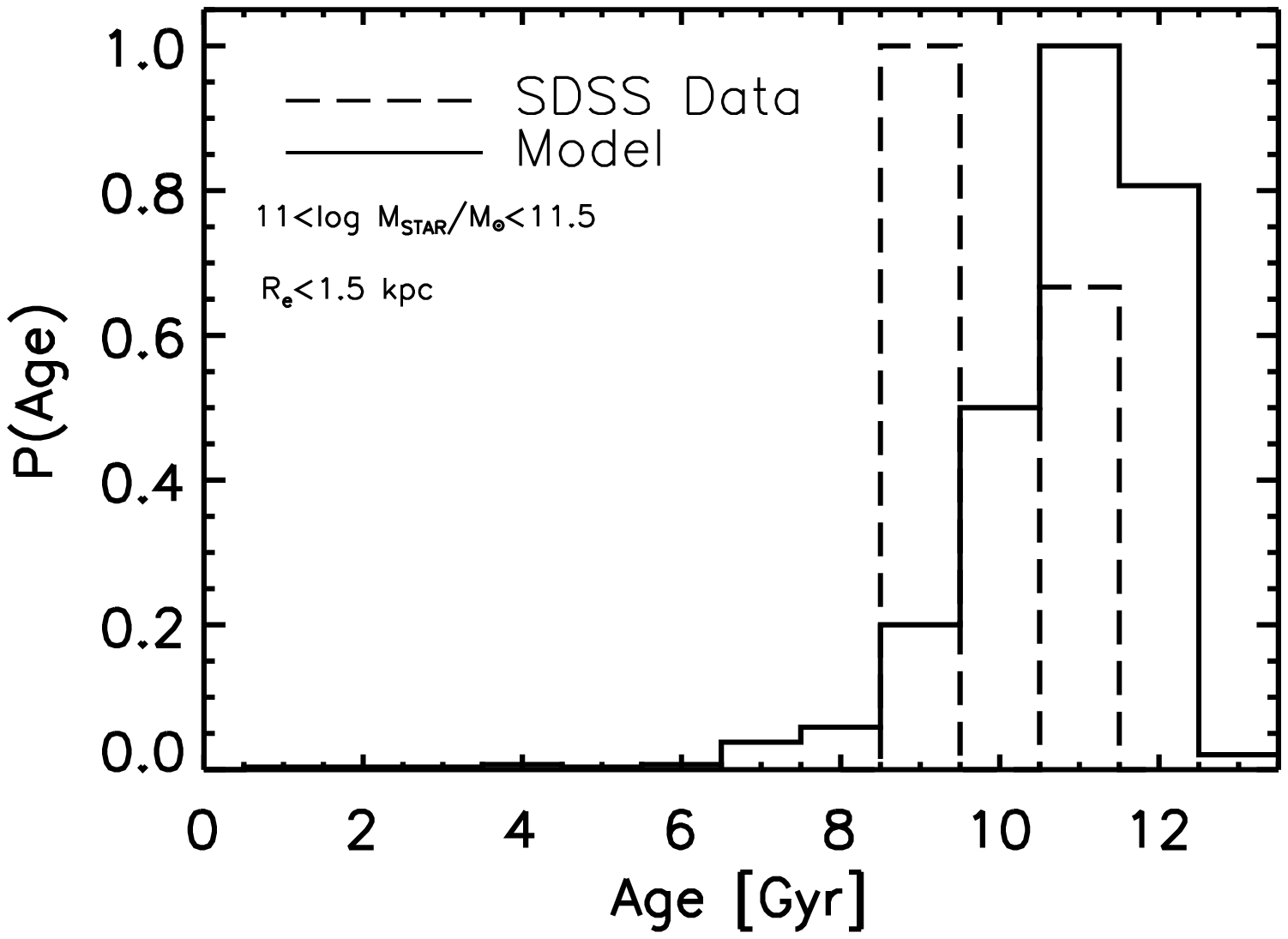}
\caption{Normalized age distribution for the spheroids and bulges
in the range of mass and size as labeled as predicted by the model
(\emph{solid} lines) and as calibrated in SDSS (\emph{long-dashed} lines). There is
a tendency for galaxies in the model to be older. This in turn might
play a role in producing too many compact galaxies
at low redshifts.}
\label{fig|Page}
\end{figure*}

First of all, we note that model galaxies tend
to be significantly older than similar
galaxies in SDSS. Figure~\ref{fig|Page} shows
normalized age distributions for model (solid lines) and
SDSS (long-dashed lines), massive and compact
galaxies, as labeled. Here the ages in SDSS are from the spectral
analysis of \citet{Gallazzi05}, and, although they might be biased
by a systematic 1.5 Gyr with respect
to mass-weighted ages (e.g., \citealt{jimenez07}), they are still systematically
lower than the predicted ones.
We recall that the analysis carried out by \citet{ShankarBernardi} and \PapI,
showed that, on statistical grounds,
the trends of age with stellar mass
and size do not significantly depend
on the exact choice for the age estimator,
although different methods might yield significantly
different ages on an object-by-object basis.
We find that while galaxies in SDSS peak at
ages around $5-8$ Gyr ($z \sim 1-1.5$) and have
a rather broad distribution with a significant number of older and
younger galaxies (see the more detailed analysis in \PapInp),
model galaxies are, on average, older, peaked around $11-12$ Gyr ($z\sim 3-5$),
and have a narrower distribution. This result
holds irrespective of the exact bulge fraction cut considered in the model
(i.e., the discrepancy is present for both S0 and pure elliptical
galaxies).
The difference in formation epoch might be, at least in part,
responsible for producing a larger number of extremely compact
galaxies at low redshifts.
The model in fact produces, at fixed stellar mass, more compact galaxies
at higher redshifts (see Figure~5 in \PapInp), as naturally expected
if older galaxies are born from gas-richer events.

\begin{figure*}
\includegraphics[width=14truecm]{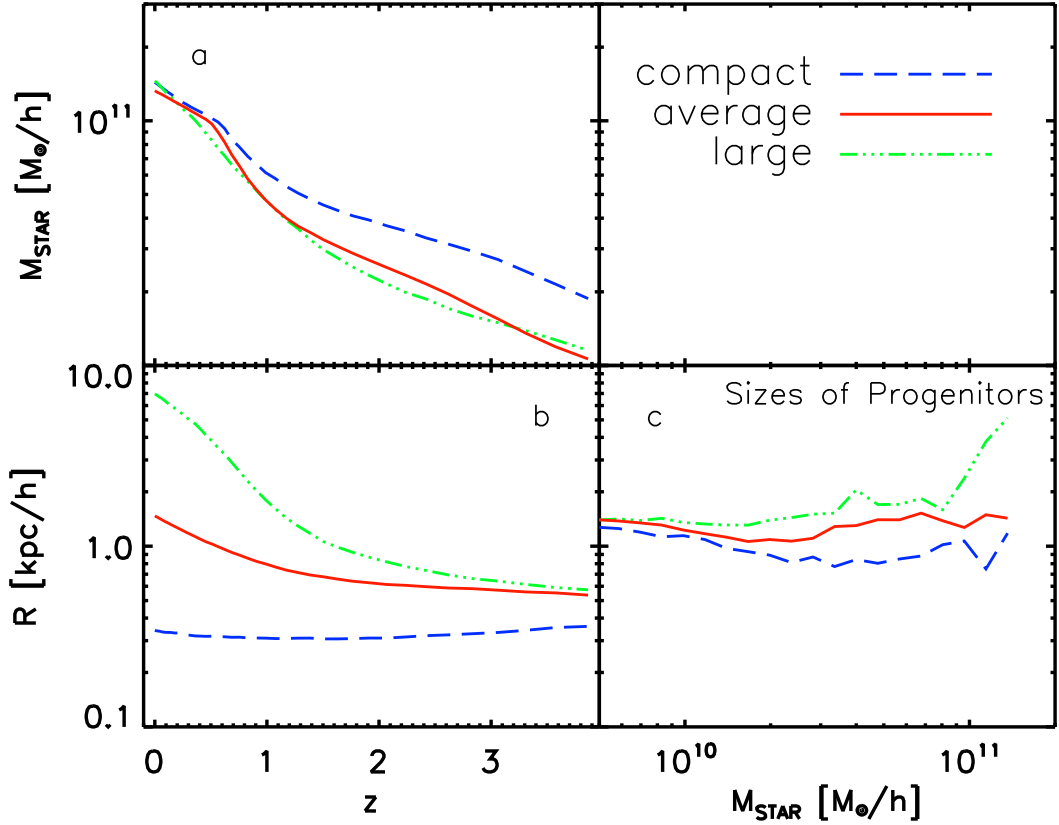}
\caption{Median \mstar\ and \re\ as a function of redshift (panels \emph{a} and \emph{b}, respectively)
for three samples of galaxies randomly extracted from the \citet{Bower06}
catalog with similar stellar mass at $z=0$ but very different final sizes (compact, large, and average, as
labeled). Panel \emph{c} shows the median radius versus stellar mass for the progenitors
of the three samples. The progenitors
of the compact, massive galaxies systematically have lower sizes.}
\label{fig|Progenitors}
\end{figure*}

To further test the origin of the suppressed
size evolution in a fraction
of massive galaxies,
we randomly select three samples of compact, average, and large bulge-dominated
galaxies from the \citet{Bower06} model at $z=0$ with stellar mass of $\sim 1.5\times 10^{11}\,$\msun, and a median
size of \re=0.4, 1.5, and 8 kpc, respectively.
Figure~\ref{fig|Progenitors} shows the comparison among the median \mstar\ and \re\ as a function of redshift (panels \emph{a} and \emph{b}, respectively) for three samples, showing that the compact galaxies start off as more massive galaxies at high redshifts but do not evolve, on average, their original sizes, even if, surprisingly, they can grow in mass by up to a factor of $\sim 10$. In Figure~\ref{fig|Progenitors}\emph{c} we plot the
overall median size-mass relation for the full set of progenitors of the galaxies
examined in panels \emph{a} and \emph{b}. The galaxies included in Figure~\ref{fig|Progenitors}\emph{c} are all the progenitors merging onto the main branch. In other words, we exclude both galaxies along the main branch and those that merge with other galaxies that will subsequently merge onto the main branch.
It is apparent that the compact galaxies
have ``progenitors'' always characterized by sizes a factor of a few smaller than average, which
we believe is one of the main causes for the negligible growth in their sizes. In fact, from the virial condition given
in Eq.~(\ref{eq|virialcondition}), it is clear that setting $\eta=M_2/M_1$ and $k=R_2/R_1$, we get
\begin{equation}
R_f=(1+\eta)^2/[1+\eta^2/k+2\eta f_{\rm orb}/(1+k)]R_1\, .
\label{eq|virialconditionEvolved}
\end{equation}
Therefore, if the merging progenitors have comparable sizes, i.e., $k\approx 1$, then this translates into
$R_f\approx (1+\eta)^2/[1+\eta^2+2\eta]= R_1$, irrespective of the value for $\eta$, while progenitors with $k>>1$ will efficiently puff-up the size of the remnant by a factor $\sim (1+\eta)^2$.
Another possible concurrent reason for the very compact remnants can be possibly associated with the choice of $f_{orb}$ in Eq.~\ref{eq|virialcondition}. Cosmological N-body simulations show in fact that merging dark matter halos mostly do so on approximately parabolic orbits \citep[e.g.,][]{KhochfarBurkert06}, implying values of $f_{orb} \approx 0$, and hence larger remnant sizes.

Overall, we conclude that, within the hierarchical framework of the \citet{Bower06} model, allowing the compact, high-$z$ galaxies to be born at more recent epochs and to possibly
merge with more massive and larger progenitors,
should in principle allow them to grow more in size
yielding better agreement with the low portion
of the observed ERF. We therefore confirm
that the discrepancy between model predictions and data regarding the presence
of compact and massive galaxies is real and it causes
the flattening of the predicted size-mass relation. Nevertheless,
the model can efficiently increase in size and mass
a large fraction of spheroids, in a way to actually
significantly overproduce the number of local, large galaxies,
at least above $\gtrsim 10$ kpc.

\section{DISCUSSION}
\label{sec|discu}

In the previous sections we showed that hierarchical
models are characterized by a two-stage evolution, a fast wet, high-redshift
phase, followed by a much longer assembly phase
dominated by minor mergers (see Figure~\ref{fig|NumberMergersFromModels}).
A fraction ($\sim 10-50\%$)
of the stellar mass of the spheroids is mainly formed through wet, high-$z$ merging events among gas-rich, disk dominated
subunits, while the rest of the stellar mass is added via later dry mergers\footnote{We stress that the growth of the bulge stellar mass via disk instability in the massive spheroids considered here is not more than 10\%, and only $\sim 2\%$ for the galaxies
with final stellar mass \mstar$\gtrsim 10^{11}$\msun.}.
This kind of evolution may be connected with the fast and slow accretion phases of single dark matter halos seen in high resolution numerical simulations \citep[e.g.,][]{Zhao03,Lu06,Diemand07,Ascasibar08,Vass09}. These studies support a scenario in which during a first, fast, and chaotic phase, the halo
builds up the central potential well, while during a second, much
longer phase, matter is accreted in a smoother way. The latter phase
could resemble a sequence of minor mergers, although the total
amount of stellar mass carried towards the central regions during
this longer phase is still unclear
\citep[e.g.,][and references therein]{Boylan08,Kaza08,Purcell08,Drory08,Perez08}. An increasing stellar mass with
time may be actually going on in disk-dominated galaxies, where the
break radius in the light profiles has been observed to increase by
a factor of 1.3 since $z\sim 1$ \citep{Azzollini08,Bakos08}.

Minor mergers might also be at the origin
of the boxy and disky
early-type dichotomy \citep[e.g.,][and references
therein]{Cappellari07,Emsellem07,Kang07,Pasquali07}. Assuming that
wet and dry mergers are actually responsible for determining the
disky and boxy nature of stellar orbits in ellipticals, then the
results in Figure~\ref{fig|NumberMergersFromModels}
would imply that older galaxies, which have undergone more
dry mergers since their formation epoch, would end up, on average,
with boxier isophotes with respect to younger galaxies (e.g., \citealt{KhochfarBurkert06,KochfarSilk06Rez}). The recent
study by \citet{Kormendy08} actually confirms that older galaxies
appear boxier. \citet{Almeida07} also discussed that several other
correlations, including the velocity dispersion, size and luminosity
are reproduced by the \citet{Bower06} model with no extra tuning of the
parameters.

During the wet phase, the correlations between the central black hole mass \mbh\ and
their host galaxy potential wells, characterized by their \sis,
might have also been settled, especially
in AGN feedback-constrained galaxy evolution models (e.g., \citealt{Granato04,Hopkins06,Monaco07}; but see also, e.g., \citealt{MiraldaKoll}).
If most of the black hole mass was already in place at the end of the wet phase, but only
about half of the host galaxy mass was assembled,
the hierarchical models considered here would then naturally imply an higher black hole mass
to stellar mass ratio with respect to the local one, i.e., a positive evolution in the
normalization of the \mbh-\mstar\ relation.
On the other hand, given that the host halo potential
well is rapidly built during the fast accretion phase,
it is also reasonable to expect the \mbh-\sis\ relation to possibly
already be fully established at the epoch of
the wet phase \citep[e.g.,][]{Granato04,Marulli08,Hop08FP}. Some
empirical works have, in fact, found only marginal evidence for
evolution in the \mbh-\sis\ relation, and possibly only in the more
massive systems, \citep[e.g.,][and references
therein]{Shields06,Gaskell09,ShankarMsigma}. However, minor dry mergers
are expected to have some impact on the initial velocity dispersion \sis\ (e.g., \citealt{CiottiReview,Naab09}). Also, late black hole re-activations (e.g., Menci et al. 2004, Vittorini et al. 2005), and/or black hole mergers (e.g., Volonteri et al. 2005, Malbon et al. 2007), might have increased the black hole masses
since the wet epoch, further influencing the evolution in the scaling
relations between black holes and their host galaxies.
Overall, the dynamical evolution of galaxies and their central black holes,
tested against the local velocity dispersion function
(Sheth et al. 2003, Bernardi et al. 2009) and fundamental plane of early-type
galaxies, should provide valuable additional insights into our understanding
of galaxy evolution.

\section{CONCLUSIONS}
\label{sec|conclu}

In this paper we make use of the
data sets derived from SDSS DR6 by
Bernardi et al. (2009) and \citet{Hyde09a}, used to derive
the size and stellar mass functions for
a sample of early-type galaxies with
concentration $C_r>2.86$, comprised of both ellipticals
and S0 galaxies, and a sample dominated by ellipticals, respectively.

We compare these statistical distributions
with the hierarchical model by \citet{Bower06}.
The aim of this exercise is to show how the
simultaneous comparison of the size and mass
distributions can reveal interesting insights
on how to improve the performance of theoretical models
of galaxy evolution.

We find, in agreement with previous studies, that
this hierarchical model provides a poor
match to the size-mass relation of local
galaxies, irrespective of the exact sample
we compare it with. In particular, the model tends to produce a
much flatter relation than the one actually observed.
This flattening is mainly produced by the combined
effects of having, with respect to the local data,
too large ($\sim 3$ kpc)
low-mass galaxies ($<10^{11}$\msun), and of having a non-negligible fraction
of compact galaxies ($\lesssim 0.5-1$ kpc) at high masses ($\gtrsim 10^{11}$\msun).

Such discrepancies are reflected in the predicted size distribution.
Although the model produces
a size distribution in broad agreement with the
data, it tends to overproduce the number of large
galaxies beyond the peak ($\gtrsim 3$ kpc),
and the number of very compact galaxies ($\lesssim 1$ kpc).
We discussed that the former issue is present at all epochs,
and it might therefore be linked to how
spheroids are formed in the first place, either from
not properly treating initial disk instabilities
and/or computing the sizes of remnants in gas-rich mergers.

Regarding the overproduction of compact
and massive (\mstar$\sim (0.5-1)\times 10^{11}$\msun) galaxies with respect to the data,
already pointed out in the recent Literature,
we find it to be less prominent than previously claimed,
and confined to only ultracompact galaxies
(\re$\lesssim 0.5$ kpc) when considering
only ellipticals.
We discuss two possible reasons behind
the survival of such compact
galaxies until the present epoch. First, we find that model
early-type galaxies tend to be significantly
older than those in SDSS. This in turn
might induce more compact galaxies at fixed stellar mass,
given that galaxies formed at higher redshifts
are more compact (see \PapInp).
We also find that model early-type
compact galaxies underwent peculiar merging histories
characterized by extremely compact progenitors,
that could prevent them to efficiently grow their sizes.

\section*{Acknowledgments}
FS acknowledges support from the Alexander von Humboldt Foundation and partial support
from NASA Grant NNG05GH77G. MB is supported by NASA grant
LTSA-NNG06GC19G and NASA ADP/NNX09AD02G. We thank Guinevere Kauffmann, Ravi Sheth, Andrew Benson, Luigi Danese, Rosalind Skelton, Simon White, Qi Guo, and Volker Springel for various discussions. We finally thank
the referee for several useful suggestions that improved
the presentation of the paper.

\bibliographystyle{mn2e}
\bibliography{../../RefMajor}

\appendix

\section{PREDICTING THE SIZE FUNCTION FROM CONVOLUTION METHODS}
\label{sec|convolutionofERF}

The filled squares in Figure~A1 represent the Bernardi et al. (2009)
estimate of the ERF obtained from the $V/V_{\rm max}$
method for our sample of early-type
galaxies selected with concentration
$C_r>2.86$.

\begin{figure*}
\includegraphics[width=8.5truecm]{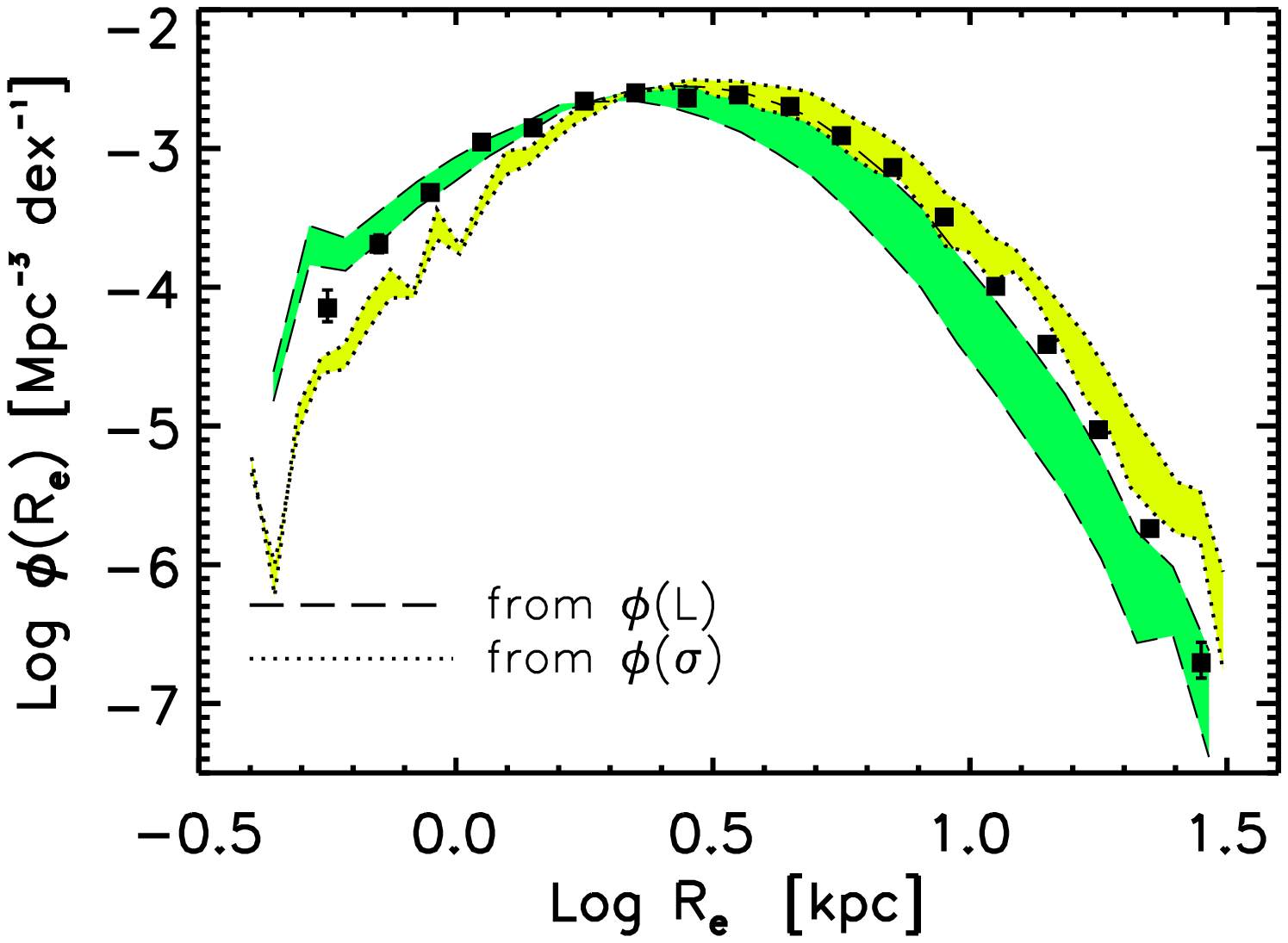}
{\caption{Size function of the $C_r>2.86$ sample obtained
from the $V/V_{\rm max}$ method (\emph{solid squares}), compared with the size
functions obtained from the convolution of the luminosity function (\emph{long-dashed} lines) and velocity dispersion function (\emph{dotted} lines), with the bivariate distributions
of points in the $L-R_e$ and $\sigma-R_e$ planes, respectively (see text).}
\label{fig|PhiRez2}}
\end{figure*}

For consistency, we here show that the $V/V_{\rm max}$-based
ERF is consistent, within the errors, with the one obtained
from the convolution of the luminosity function or velocity dispersion
function with the bivariate distribution of points in
the $L-R_e$ plane.
More specifically, following the methods outlined
in \citet{Sheth03} and \citet{Shankar04},
we have convolved the luminosity function $\Phi(L)$
with the bivariate distribution of $L_j$ and $R_e$,
\begin{equation}
\Phi(R_e^i)=\sum_{j}\xi_{ij}\Phi(L_j)\, .
    \label{eq|bivar}
\end{equation}
Here $\xi_{ij}$ is the fraction of sources in the sample with
effective radius $R_e^i$ and luminosity $L_j$, normalized to
the total number of sources with luminosity $L_j$.
The luminosity function $\Phi(L)$ has been
computed from the $V/V_{\rm max}$ method
by Bernardi et al. (2009) for the same sample
of galaxies, and
we refer the reader to that paper
for analytical fits and detailed
discussions of the sample.
The result of Eq.~(\ref{eq|bivar}) is shown in Figure~A1
with long-dashed lines, which bracket the statistical uncertainties
in the luminosity function fit parameters.
We have also used the bivariate distribution in equation~(\ref{eq|bivar})
applied to the velocity
dispersion function $\Phi(\sigma)$, again derived by
Bernardi et al. (2009) for this same sample, and with the weights
$\xi_{ij}$ now computed from the distribution
of sources in the $\sigma-R_e$ plane. The result
is shown with dotted lines in the same Figure, again
bracketing the statistical uncertainties in the velocity
dispersion function fit parameters.

This exercise proves that, as expected,
convolutions of other statistical distributions $\Phi(x)$
with their appropriate scaling relations $x-R_e$, provide
consistent results. However, it also shows that the accuracy of the
results relies on the accuracy of the
input weights $\Phi(x)$, that in turn proves
the importance of directly adopting the $V/V_{\rm max}$ method to derive a
more precise estimate of the ERF.

\label{lastpage}

\end{document}